\documentclass[prd,nofootinbib,preprintnumbers,superscriptaddress]{revtex4}

\usepackage{amsmath,amssymb}
\usepackage{graphicx}
\usepackage{url}
\usepackage{color}

\newcommand{\mpl}{\ensuremath{M_\text{Planck}}}
\newcommand{\mgut}{\ensuremath{M_\text{GUT}}}
\newcommand{\mess}{\ensuremath{M_\text{mess}}}
\newcommand{\med}{\ensuremath{M}}

\newcommand{\sq}[1]{\ensuremath{\tilde{q}_{#1}}}
\newcommand{\squ}[1]{\ensuremath{\tilde{u}_{#1}}}
\newcommand{\sqd}[1]{\ensuremath{\tilde{d}_{#1}}}

\newcommand\one{\leavevmode\hbox{\small1\normalsize\kern-.33em1}}

\newcommand{\ord}{\mathcal{O}}
\newcommand{\qqquad}{\qquad \qquad}

\newcommand{\go}{\ensuremath{\tilde{g}}}
\newcommand{\sqb}[1]{\ensuremath{\tilde{b}_{#1}}}

\newcommand{\nz}[1]{\ensuremath{\tilde{\chi}_{#1}^0}}
\newcommand{\cpm}[1]{\ensuremath{\tilde{\chi}_{#1}^\pm}}

\newcommand{\mgo}{\ensuremath{M_{\tilde{g}}}}
\newcommand{\msq}[1]{\ensuremath{m_{\tilde{q}_{#1}}}}

\newcommand{\mse}[1]{\ensuremath{m_{\tilde{e}_{#1}}}}

\newcommand{\mch}[1]{\ensuremath{m_{\tilde{\chi}^+_{#1}}}}
\newcommand{\mne}[1]{\ensuremath{m_{\tilde{\chi}^0_{#1}}}}

\newcommand{\ifb}{{\ensuremath\rm fb^{-1}}}

\def\slashchar#1{\setbox0=\hbox{$#1$}           
   \dimen0=\wd0                                 
   \setbox1=\hbox{/} \dimen1=\wd1               
   \ifdim\dimen0>\dimen1                        
      \rlap{\hbox to \dimen0{\hfil/\hfil}}      
      #1                                        
   \else                                        
      \rlap{\hbox to \dimen1{\hfil$#1$\hfil}}   
      /                                         
   \fi}

\def\eg{{\sl e.g.} \,}
\def\ie{{\sl i.e.} \,}

\begin{document}
\numberwithin{equation}{section}

\title{Travels on the squark-gluino mass plane}

\author{Joerg Jaeckel}
\affiliation{Institute for Particle Physics Phenomenology,
             Department of Physics,
             Durham University, United Kingdom}

\author{Valentin V. Khoze}
\affiliation{Institute for Particle Physics Phenomenology,
             Department of Physics,
             Durham University, United Kingdom}

\author{Tilman Plehn}
\affiliation{Institut f\"ur Theoretische Physik,
             Universit\"at Heidelberg,
             Germany}

\author{Peter Richardson}
\affiliation{Institute for Particle Physics Phenomenology,
             Department of Physics,
             Durham University, United Kingdom}
\date{\today}

\preprint{IPPP/11/52; DCPT/11/104}

\begin{abstract}
\noindent
Soft supersymmetry breaking appears in the weak-scale effective action
but is usually generated at higher scales. For these models the structure of the
renormalization group evolution down to the electroweak scale 
leaves only part of the squark-gluino and
slepton-gaugino mass planes accessible.  Our observations divide
these physical mass planes into three wedges: the first can be reached
by all models of high-scale breaking; the second can only be populated
by models with a low mediation scale; in the third wedge squarks and
gluinos would have to be described by an exotic theory. All usual
benchmark points reside in the first wedge, even though an LHC
discovery in the third wedge would arguably be the most exciting
outcome.
\end{abstract}

\maketitle

\newpage

\section{Introduction}
\label{sec:intro}

Searches for supersymmetry are one of the most visible tasks of the
LHC experiments~\cite{atlas,cms}. To interpret the data they 
have to rely on specific SUSY models determining the mass spectrum and
the decay patterns.
Limiting the Higgs sector to two doublets, a good starting point for such
an interpretation is the MSSM defined at the weak
scale. However, for practical purposes one needs to significantly
constrain its vast parameter space. After taking into account the
strong constraints for example from flavor physics~\cite{flavor} and
electric dipole moments~\cite{edm} we are left with $\ord(20)$
parameters which can be relevant for LHC searches or
observations~\cite{sfitter}. A further reduction of this parameter
space is traditionally achieved in terms of simplest constrained
realizations such as the CMSSM/mSUGRA~\cite{Chamseddine:1982jx,Tata:1997uf}, gauge 
mediation~\cite{Dine:1981gu,Meade:2008wd,Giudice:1998bp} or anomaly
mediation~\cite{Randall:1998uk} (see also \cite{primer} for an overview).

These models share two important features. First, by construction they
have a small or even minimal number of free parameters to describe all
the soft supersymmetry breaking terms.  Second, the soft parameters in
these models are determined at a high scale arising from an underlying
theory of supersymmetry breaking and mediation. At the mediation scale
$\med$ the values for the soft parameters are initialized.  For
example, in gauge mediation, $\med = \mess$, an effective mass of the
messenger fields transmitting supersymmetry breaking to the Standard
Model sector. In these models $\mess$ is typically taken to be in the
range $10^{5}-10^{14}$~GeV.  In gravity mediation models $\med$ is set
by $\mpl$ which, with the additional assumption of grand unification,
in CMSSM is traded down to $\mgut$.  In order to make contact with the
scale at which experiments operate the soft terms have to undergo
renormalization group evolution down from the mediation scale $\med$
to the weak scale~\cite{primer,susy_rge}.\bigskip

In this paper we point out that all such high scale models automatically impose
severe restrictions on superpartner masses at collider energies.  In
the strongly as well as weakly interacting sfermion-gaugino mass
planes as much as half of the available parameter space becomes
inaccessible. For example in the squark-gluino case, squarks cannot
become significantly lighter than gluinos. Similar relations hold
between sleptons and electroweak gauginos in all high scale models.

The details of these restrictions dominantly depend on one parameter,
namely the value of the mediation scale.  As a result, each
sfermion-gaugino mass plane can be divided into three wedge-shaped
regions. One region can be reached by all usual models of high-scale
supersymmetry breaking. A second wedge can only be populated by models
with a mediation scale $\med < \mgut$, while sfermions and gauginos
in the third wedge would have to originate from a theory which either
does not have a high SUSY scale or a qualitatively different RG
evolution.  Thus, from measuring gaugino and sfermion masses we can 
draw powerful conclusions on the way supersymmetry
is realized in Nature.

Conversely, when searching for supersymmetry one should make as few
assumptions as possible about the way supersymmetry breaking is
realized. This definitely includes its high scale origin. 
With the next round of SUSY searches at the
LHC being imminent, new sets of benchmark points and test models will
be defined to determine, optimize and calibrate the search strategies.
In order to minimize the bias of assumed specific models 
it may be useful to include also points which do not originate
from high scale models and which are distributed more democratically
on the sfermion-gaugino planes accessible at collider energies.  One
way to obtain points not prejudiced towards high scale models is to
use the MSSM defined directly at the weak scale.  A manageable
incarnation of this idea is the so-called phenomenological MSSM or
pMSSM~\cite{pmssm}.  Alternatively, one can use the simplified model approach for
constructing test models based on kinematic considerations and a
selection of a small number of allowed sparticle
species~\cite{simplified}. To some degree, squark and
gluino searches both by ATLAS~\cite{atlas} and CMS~\cite{cms} are already following this route.\bigskip

This paper is organized as follows: in the next section we
will show explicitly how the renormalization group evolution from the
high scale $\med$ to the weak scale restricts the accessible regions
in the squark-gluino plane. In Sec.~\ref{sec:msl} we extend our
discussion to binos, winos and sleptons. In particular, we discuss the additional
complications arising from electroweak symmetry breaking.  In
Sec.~\ref{sec:bench} we investigate the distribution of benchmark points
as well as a variety of test models. Finally, in
Sec.~\ref{sec:conclusions} we summarize our findings and conclude.
 
\section{Squark vs gluino mass}
\label{sec:msq}

The key to understanding the coverage of the squark-gluino mass plane
is the renormalization group equation for these masses. It has been
known for a long time~\cite{primer,susy_rge} that the gaugino masses strongly
impact the running of the sfermion masses to the weak scale.  Starting
from a high scale, they generate contributions to the soft sfermion
masses even if the initial soft sfermion masses vanish. 

To illustrate this structure, we can approximately solve the
RG equations in SUSY-QCD. In the absence of Yukawa
couplings we find schematically
\begin{alignat}{5}
\msq{}^2(Q) &\sim \msq{}^2(\med) 
\, + \, A_{\sq{}} \msq{}^2
\, + \, A_{\go} \mgo^2 \notag \\
\mgo(Q) &\sim \mgo(\med)
\, + \, B_{\go} \mgo \; .
\end{alignat}
Numerically, $A_{\go}$ dominates. The running of the gluino mass does
not include any squark mass terms on the right-hand side. The reason
is that Majorana fermion masses are protected by the $R$ symmetry, in
analogy to the chiral symmetry for the Dirac masses of the Standard
Model fermions. This feature persists for the entire MSSM and can be
exploited for example to decouple all scalars from a high-scale SUSY
model while keeping all gauginos light enough to ensure gauge coupling
unification, dark matter, etc~\cite{split_susy}.\bigskip

Moving on to the full theory, for the first two generations we neglect
the Yukawa couplings~\cite{flavor}.  If the trilinear $A$-terms are
proportional to the Yukawa couplings, the same holds for the
renormalization group contributions from them. Using this, the RG
equations for the first generation sfermions read~\cite{primer},
\begin{equation}
16\pi^2\frac{d}{dt} m_{\tilde{f}}^2 =
- 8 \sum_r C_r g_r^2 |M_r|^2 
\, + \, 2 Y_{\tilde{f}} g_1^2 S \; ,
\label{eq:scalar}
\end{equation} 
where $M_r$ are the gaugino masses, $r=(1,2,3)$ the (bino,wino,gluino)
labels, $g_r$ the gauge couplings not in the GUT normalization for
$U(1)$, and
\begin{equation}
S:={\rm Tr}(Ym^2)
=\sum_\text{generations}
 \left( m_{\tilde{Q}_L}^2 
     - 2m_{\tilde{u}_R}^2
     +  m_{\tilde{d}_R}^2 
     -  m_{\tilde{L}_L}^2 
     +m_{\tilde{e}_R}^2 \right)
 +m^2_{H_{u}}-m^2_{H_{d}} \; .
\end{equation}
The Casimir invariants and hypercharge assignments for the relevant
fermions are
\begin{alignat}{5}
(C_1, C_2, C_3) &=
\left( Y^2, \frac{3}{4}, \frac{4}{3} \right) \notag \\
(Y_{\tilde{Q}_L},Y_{\tilde{u}_R},Y_{\tilde{d}_R},Y_{\tilde{L}_L},Y_{\tilde{e}_R}) &=
\left( \frac{1}{6},-\frac{2}{3}, \frac{1}{3}, -\frac{1}{2}, 1  \right) \; .
\end{alignat}
The gaugino masses, couplings, and scalar masses then evolve according to 
\begin{alignat}{5}
16\pi^2 \frac{d}{dt} M_r
&= - 2 b_r g_r^2 M_r \notag \\
16\pi^2 \frac{d}{dt} g_r^2
&= - 2 b_r g_r^4 \notag \\
16\pi^2 \frac{d}{dt} S
&= - 2 b_1 g_1^2 S
\qqquad (b_1,b_2,b_3) = \left( -11, -1, 3 \right).
\label{eq:ino}
\end{alignat}
Comparing Eq.\eqref{eq:scalar} and Eq.\eqref{eq:ino} we indeed see
that the gaugino masses contribute to the running of the sfermion
masses but not vice versa.\bigskip

Equation~\eqref{eq:scalar} can easily be integrated,
\begin{alignat}{5}
m^2_{\tilde{f}}(Q) =
m^2_{\tilde{f}}(\med) 
\, + \, \frac{Y_{\tilde{f}}}{b_1}
  \left[ \frac{\alpha_1(Q)}{\alpha_1(\med)}-1 \right] \, S(\med)
\, + \, \sum^3_{r=1} \, \frac{2 C^{\tilde{f}}_r}{b_r} \,
  \left[ 1 - \frac{\alpha_r^2(\med)}{\alpha_r^2(Q)} \right] \; M^2_r(Q) \; .
\label{eq:slope}
\end{alignat}
Because we will mainly be interested in sfermion masses smaller than the
gaugino masses, the on-shell corrections to the gaugino masses are
small, so we can identify the gaugino mass parameter with the mass
$M_r$. In addition, we can average over the light squark masses.  The
term proportional to the hypercharge and $S$ then drops out and we
find for the average squark mass
\begin{alignat}{5}
m^2_{\sq{}}(Q):\,=\, \frac{1}{4}\left[2m^2_{\tilde{Q}_L}+m^2_{\tilde{u}_R}+m^2_{\tilde{d}_R}\right](Q) 
\,=\, m^2_{\sq{}}(\med)  
 + \, \frac{1}{4}\,\sum_{\tilde{f}=2\tilde{Q}_L,\tilde{u}_R,\tilde{d}_R}
                   \sum^3_{r=1} \, \frac{2 C^{\tilde{f}}_r}{b_r} \,
\left[1-\frac{\alpha_r^2(\med)}{\alpha_r^2(Q)} \right] 
\; M_r^2(Q) \; .
\label{eq:slope2}
\end{alignat}
Similar averaged expressions can be obtained for the
sleptons.\bigskip

At scales $Q$ below $\med$ the $U(1)$ and $SU(2)$ running implies
$\alpha(\med)/\alpha(Q)>1$, while for $SU(3)$ this ratio is less than
one. Thus all three terms in the $r$ sum on the right hand side of
Eq.\eqref{eq:slope2} are positive.  Assuming that the initial soft
sfermion mass terms are non-negative, \ie avoiding tachyonic sfermions
at the high scale\footnote{For models with high scale tachyons see~\cite{tachyons}.}, we obtain minimal sfermion mass
values at the low scale $Q$ as a function of the gaugino masses.

\begin{figure}[t]
\includegraphics[width=0.39\textwidth]{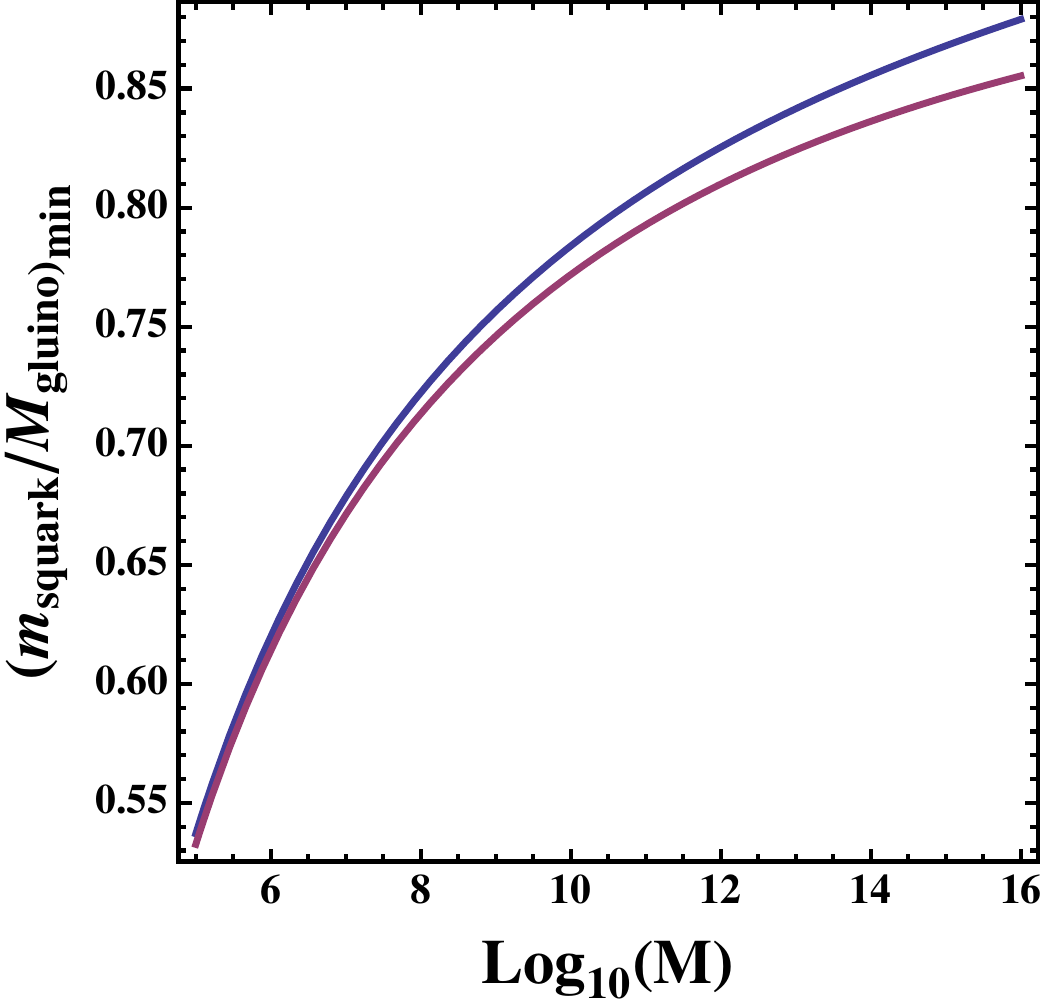}
\hspace*{0.1\textwidth}
\includegraphics[width=0.40\textwidth]{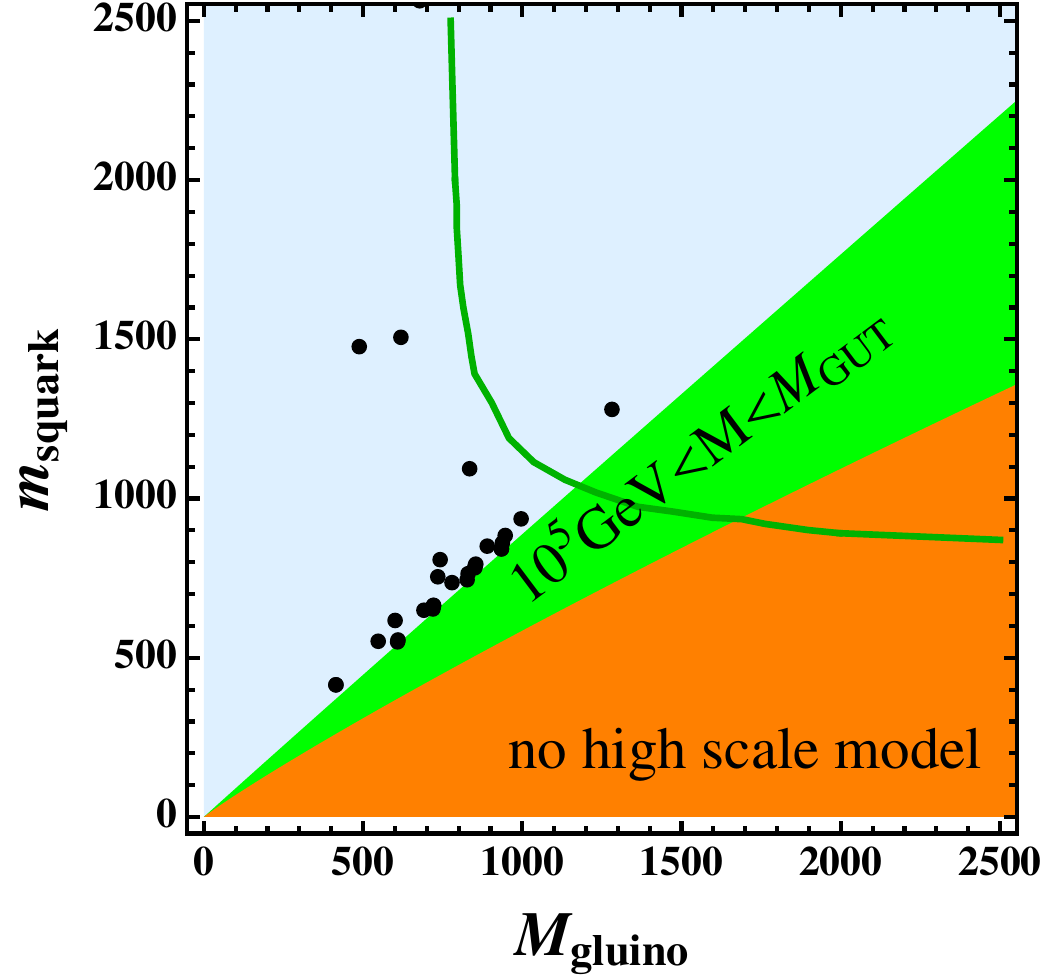}
\caption{Left: minimal ratio $\msq{}/\mgo$ as a function of the
  mediation scale. The blue curve assumes universal gaugino masses at
  the GUT scale, whereas for the red curve only $\mgo(\mgut)$ is
  non-zero. Right: accessible regions in the $\msq{}$-$\mgo$ plane,
  assuming gaugino mass unification.  Their boundaries correspond to
  mediation scales $\med =\mgut=2\cdot 10^{16}$~GeV and $\med =10^5$~GeV. The
  thick green line shows the simplified model ATLAS exclusion with
  $1.04~\ifb$~\cite{ATLASsimpl}. The dots show benchmark points from
  Refs.~\cite{atlas,cms,sps,adjk2}.}
\label{fig:sq_go}
\end{figure}

Our argument is most straightforward for the first generation squarks
where electroweak symmetry breaking effects play no role.  Given a fixed gluino mass we find the
lowest possible squark mass when the wino and bino masses vanish.
The red curve in the left panel of
Fig.~\ref{fig:sq_go} gives the minimal ratio of squark to gluino mass
averaged over $\squ{L,R}$ and $\sqd{L,R}$.  If instead of very light
weak gaugino masses we assume gaugino unification this mass ratio
slightly increases, as can be seen from the blue curve in Fig.~\ref{fig:sq_go}.

Different mediation scales, which we implicitly assume for any SUSY
model, put restrictions on the achievable physical squark masses in
terms of a lower limit on $\msq{}/ \mgo$.  The constraints on the
mass-ratio $\msq{}/ \mgo$ can be interpreted as region boundaries on
the two-dimensional squark-gluino mass plane as shown on the right
panel in Fig.~\ref{fig:sq_go}.

Beyond our basic observation, we need to make a technical aside on the
role of the low scale in Fig.~\ref{fig:sq_go}.  Eq.\eqref{eq:slope2}
depends on the choice of the renormalization point $Q$ defining
the physical masses observable at the LHC. This dependence is
logarithmic and therefore quite weak.  In the left panel of
Fig.~\ref{fig:sq_go} we simply choose $Q=1$~TeV.  In the right panel
we included this dependence by evaluating $\msq{}(\msq{})$ and
$\mgo(\mgo)$. Therefore, the lines separating the three regions are
not entirely straight.

Ignoring any high-scale physics features we start from
phenomenological weak-scale SUSY models populating the entire
squark-gluino mass plane.  The more we then increase the scale of
mediation, the stronger the constraints become and the smaller the
area in the mass plane we can cover.  Turning this argument around,
the position of a low-energy supersymmetric model on the squark-gluino
plane can be used to find an upper limit on the possible mediations scales or even make a
statement about the absence of such a scale.

In the right panel of Fig.~\ref{fig:sq_go} we divide the full plane
into three regions: region~I (blue) can easily be reached by all known
models of SUSY-breaking, including gravity and gauge mediation. 
To illustrate this we have also indicated in Fig.~\ref{fig:sq_go} a set of
SUSY benchmark points proposed and studied over the years
in Refs.~\cite{atlas,cms,sps,adjk2}. 
Region~II (green) corresponds to SUSY models where the breaking is
mediated in the window $10^5\,{\rm GeV} - M_{\rm GUT}$. It is not accessible to
gravity mediation but provides a good home for gauge mediation. 
Finally, if SUSY should be discovered
in Region~III (orange) its breaking would have to be described by an
exotic theory. It would have to descend from a theory with no or
little separation between the electroweak and the SUSY mediation
scales, excluding anything similar to gauge and gravity
mediation. These and other possibilities will be further discussed in Sect.~\ref{sec:conclusions}.\bigskip

There are different ways to study region~III. One way is to start from the 
so-called phenomenological MSSM (pMSSM)~\cite{pmssm}
Lagrangian where all MSSM soft parameters are defined at the weak scale and
no assumptions on the SUSY breaking mechanism need to be made. For studies along these lines
see~\cite{sfitter}. 
Alternatively we can utilize the
simplified model approach~\cite{simplified}, where one reduces the
number of decay topologies and with it the parameter dependence of
branching ratios to a level where only the masses of the particles
appearing in the production and decay channels have to be
tracked. 
The main difference between these two approaches is that for the weak-scale
pMSSM several decay topologies can contribute to a given signature and
that non-trivial branching ratios are included in the analysis. 
From
our point of view both approaches are well suited to avoid LHC
searches based on a theory bias.

One example is a simplified model with light squarks and gluinos and a
massless neutralino. The 95~\% confidence level exclusion contour for
this model based on $1.04~\ifb$ of ATLAS data from~\cite{ATLASsimpl} is shown in
Fig.~\ref{fig:sq_go}.  The considerable mismatch between these
exclusion contours and their CMSSM counterparts~\cite{atlas,cms}
(where they are defined) is mostly due to different neutralino mass
assumptions. We can, however, easily modify the simplified model by
assuming a light rather than massless bino and locking its mass to an
appropriately rescaled gluino mass, $\mne{1} \sim M_1 =
\alpha_1/\alpha_3 \, M_3$. This would largely be equivalent to the
high-scale motivated models where they are possible, while avoiding
assumptions about the scale and the precise nature of SUSY mediation
mechanisms which should be results of an analysis instead of
assumptions.\bigskip

Last but not least, it should be noted that the regions of the
squark-gluino plane which lie outside the usual high scale motivated
region are particularly interesting from a phenomenological point of
view. If a gluino (or other color octet) becomes significantly heavier
than the color-triplet squark it is likely that we will reconstruct
two hard decay jets, in addition to the well understood softer QCD jet
radiation~\cite{jet_rad}. The observation of for example four such
hard jets would clearly point to the production of a pair of color
octet particles~\cite{autofocus}. The reconstruction of the effective
mass is also easier if we see several hard decay products, so we can
correlate it with the number of jets, to get a first global guess at
the properties of the new particles~\cite{autofocus}. Finally, longer
on-shell decay chains with hard decay products are the basis of any
kind of SUSY parameter analysis, which for example rely on the decay
$\go \to \sqb{1} \to \nz{2} \to \tilde{\ell} \to
\nz{1}$~\cite{osland,sfitter}.

\section{Slepton vs bino/wino mass}
\label{sec:msl}

Similarly to the squark-gluino mass plane discussed in
Sect.~\ref{sec:msq}, we can also project SUSY models onto the
electroweak slepton-gauginos mass planes. Again, the region attainable
for models with a reasonably high mediation scale turns out to be
wedge shaped.

\begin{figure}[t]
\includegraphics[width=0.38\textwidth]{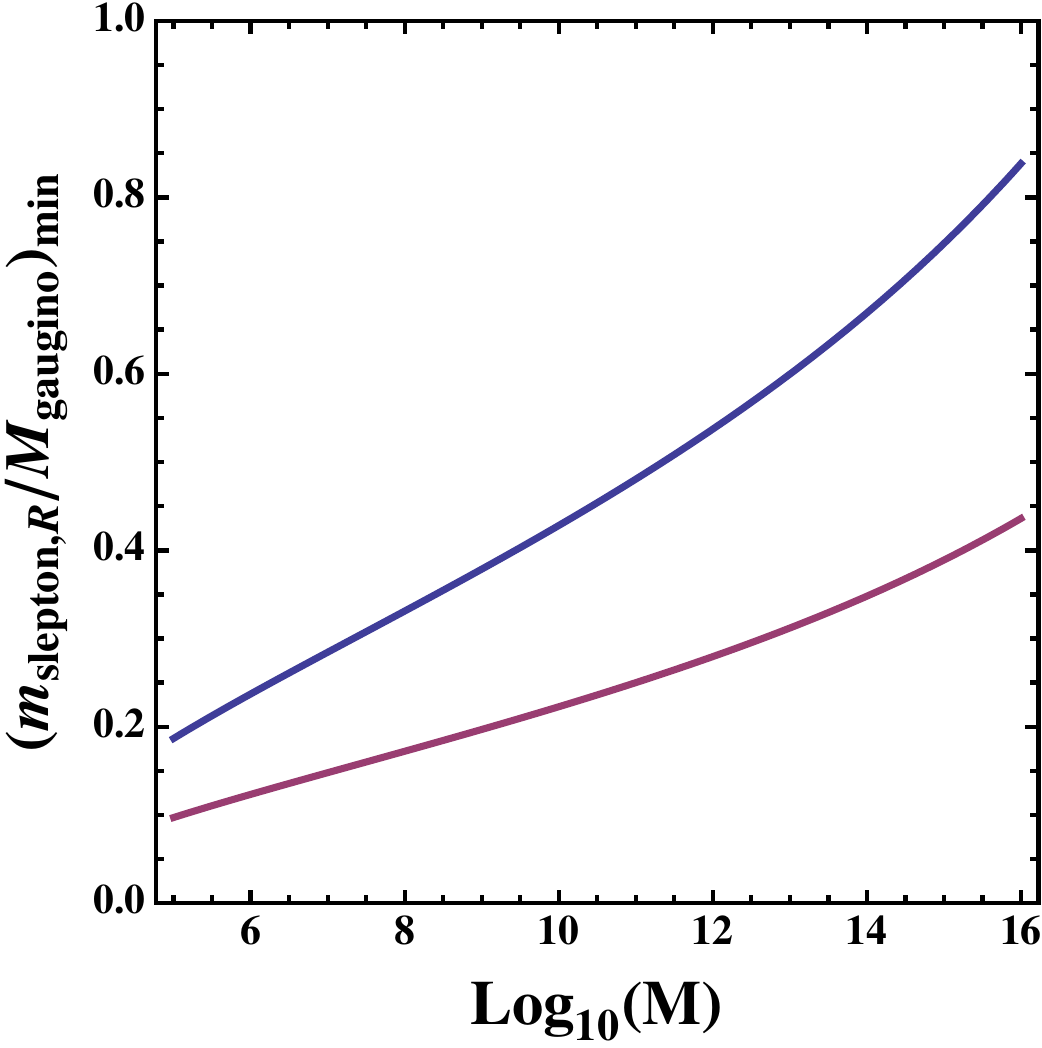}
\hspace{0.1\textwidth}
\includegraphics[width=0.38\textwidth]{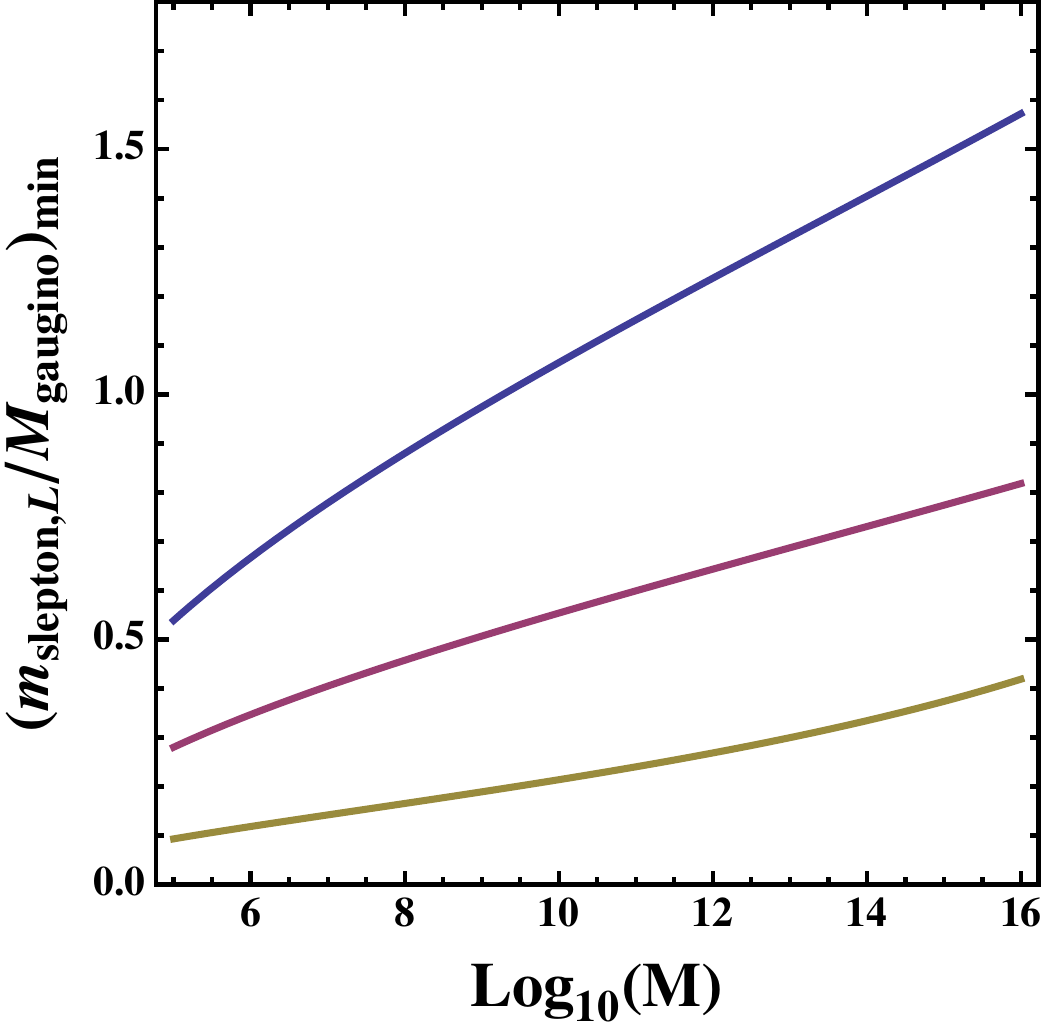}
\caption{Left: minimal ratios $\mse{R}/M_j$ for the bino (blue) and
  wino (red) as a function of the messenger scale.  Whereas the bino
  curve is independent of all other masses, the wino curve assumes
  universal gaugino masses. Right: minimal ratios $\mse{L}/M_j$ for
  bino/wino (blue/red), assuming universal gaugino masses. For the
  yellow curve which shows $\mse{L}/M_1$, only the bino mass is taken to be non-zero.}
\label{fig:mse_slope}
\end{figure}

\begin{figure}[b]
\includegraphics[width=0.40\textwidth]{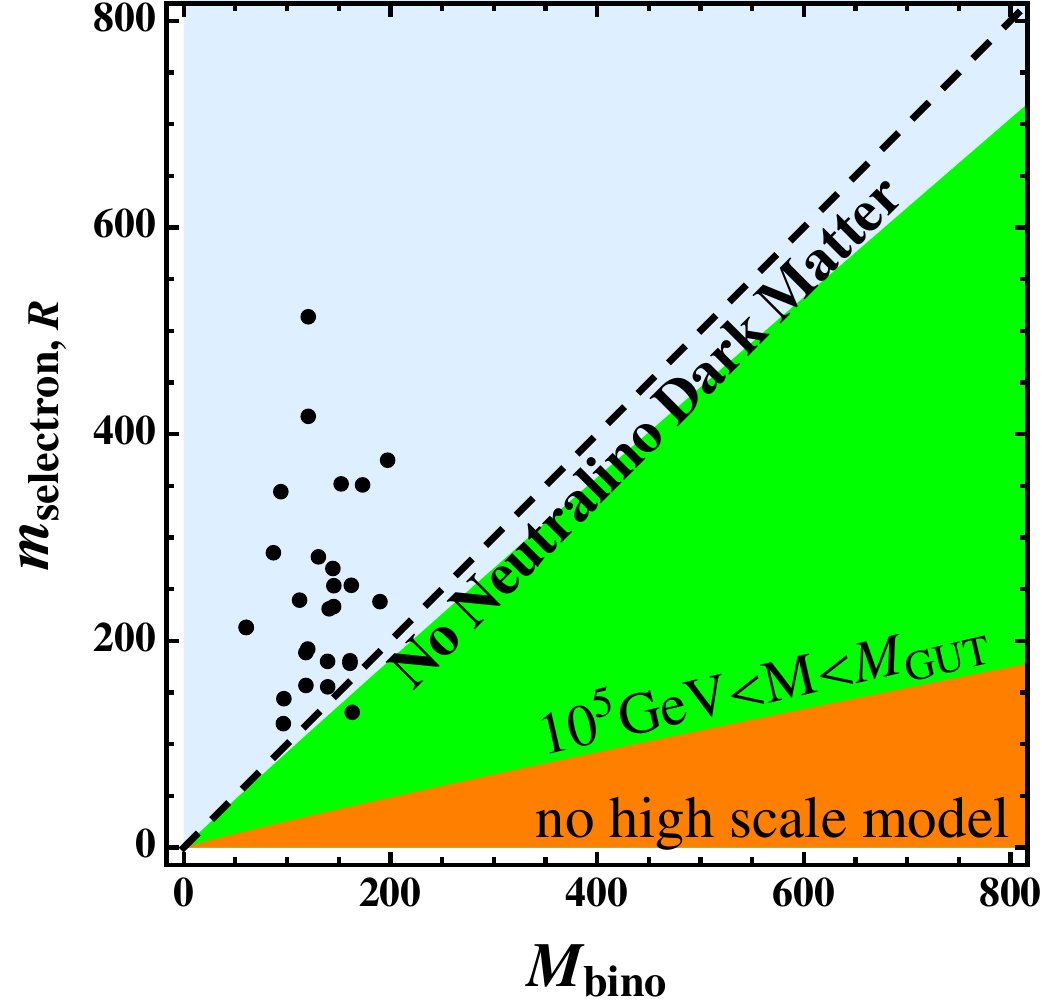}
\hspace{0.1\textwidth}
\includegraphics[width=0.40\textwidth]{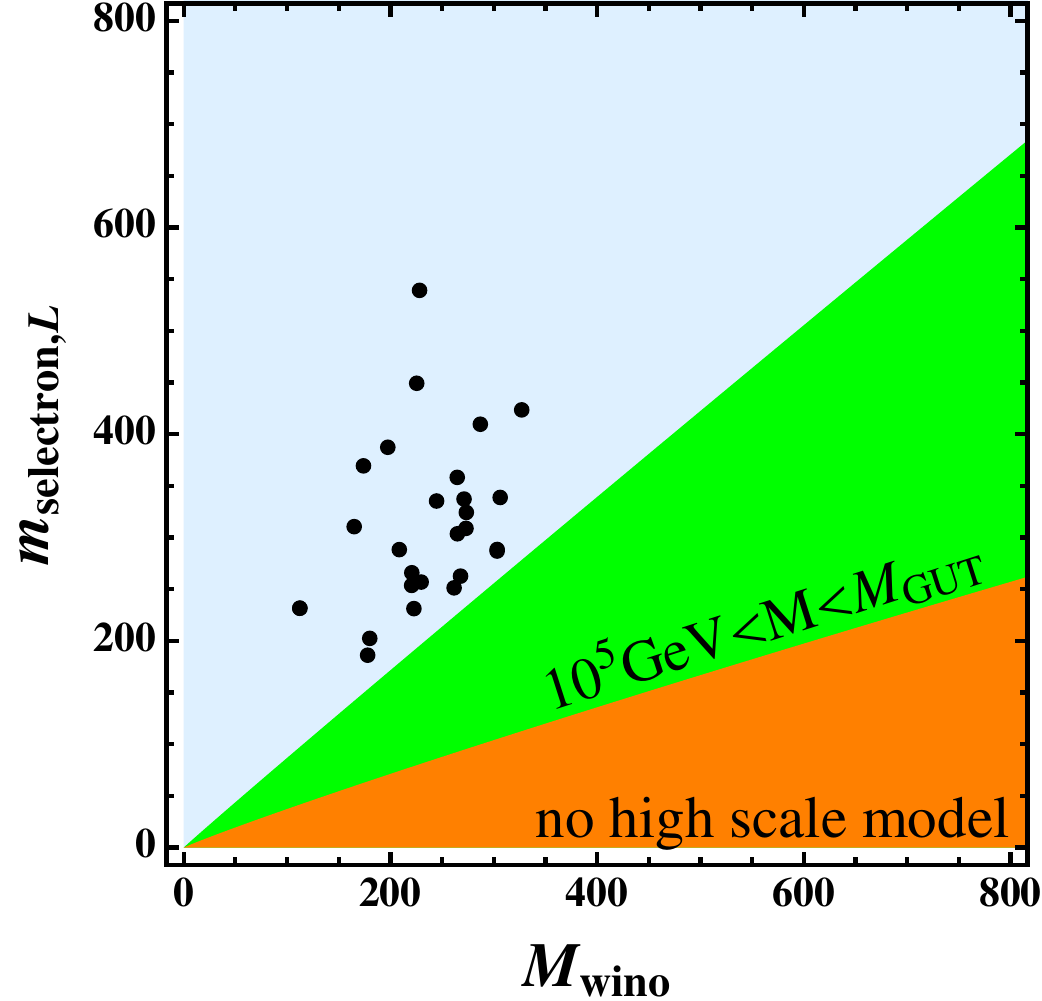}
\caption{Three regions in the sfermion-gaugino mass plane, for a bino
  or wino and left and right handed selectrons. The color coding is the same
  as in in Fig.~\ref{fig:sq_go}. We assume gaugino mass unification.
  The ``No Neutralino Dark Matter'' diagonal indicates where selectrons are
  lighter than the lightest neutralino. 
  We also display benchmark points presented in~\cite{atlas,cms,sps,adjk2}.
  In the left panel the dots indicate $\chi^{0}_{1}$ and selectron masses whereas in the right panel they
  correspond to $\chi^{\pm}_{1}$ and left handed selectron masses.}
\label{fig:mse_plane}
\end{figure}

Using Eq.\eqref{eq:slope} we can compute ratios of the left and right handed 
slepton masses to the bino and wino masses.  
Before doing that let us address the term $\propto S$ which is not positive definite.
In the simplest and most commonly used models, like the CMSSM or gauge mediation 
with universal Higgs masses,
$S=0$ at the mediation scale and remains so 1-loop. Thus the second term in Eq.~\eqref{eq:slope} is absent.
For more general models, including models with non-universal Higgs masses the $S$ term is generally non-zero.
One way to address this issue would be to average over the charged sleptons similarly to what was done for squarks in the previous section.
Instead we will choose to work with left and right handed sleptons separately and make use of the fact that the hypercharge
has opposite sign for the left and the right handed
species. Therefore, if the effect is to lower the sfermion to gaugino mass ratio in one case, it will 
unavoidably increase it in the other.
Thus we will proceed with the determination of the minimal slepton to gaugino mass ratios derived from Eq.~\eqref{eq:slope}
{\emph{without}} the $S$ term. The caveat is that a non-zero $S$ has the potential to lower either the right or left handed sfermion masses
but never both. Hence one of the minimal ratios cannot be lowered. 

We show the minimum values for all four combinations of left and right handed sleptons compared to bino and wino in Fig.~\ref{fig:mse_slope}.
The
corresponding regions in the two-dimensional mass planes are shown in
Fig.~\ref{fig:mse_plane}.  Naively, one would think that the
renormalization group running should be flatter than in the case of squarks, due to the smaller
gauge couplings. However, the relative contribution of the gauginos to
the sfermion masses in Eq.\eqref{eq:slope} is proportional to the
relative change in the gauge coupling divided by the beta function
coefficient which is of the same order of magnitude for all three
gauge groups.

The reason for the significant difference between the ratios for the
left and right handed selectrons is the chiral nature of the electroweak
interactions and the gauge structure of the gauginos. Even if the
initial supersymmetry breaking for the sfermions is chirality blind,
left and right handed sfermions will be split during the
renormalization group evolution. The typical example for such a
mediation is gravity. In contrast, gauge mediation does have a chiral
structure already at the messenger scale.\bigskip

The main difference between the squark-gluino case and the
slepton-gaugino results shown in Fig.~\ref{fig:mse_plane} is the
translation of the Lagrangian parameters into the masses of the
physical states. While for the gluino we only have to take into
account a moderately small correction to the on-shell mass scheme, the
weak gauginos are generally mixed.

What we can study at the Lagrangian parameter level, however, are the
different slepton masses. From Fig.~\ref{fig:mse_slope} we already
know that renormalization scale evolution separates left and right handed
selectrons.  For universal gaugino masses the contribution to the left
handed sfermions is always bigger and therefore left handed sfermions
are heavier. 

The two-dimensional slepton mass plane in Fig.~\ref{fig:ordering}
shows the ordering of the left and right handed masses as a function
of the bino and wino masses assuming chiral degeneracy at the
messenger scale. Gaugino mass unification, as often assumed in LHC
searches, implies $M_2 \sim 2 M_1$. This translates into a solid
prediction $\mse{L} > \mse{R}$.  However, for non-universal gaugino
masses~\cite{how_light} this can be different.  If the bino is
significantly heavier than the wino the right handed sfermions could
indeed be heavier.  Therefore, in the same way that we should not
unnecessarily assume the squark-gluino mass hierarchy as described in
the previous section, LHC searches should not be based on the
assumption that the lighter sleptons do not couple to the
wino.\bigskip

\begin{figure}[t]
\includegraphics[width=0.40\textwidth]{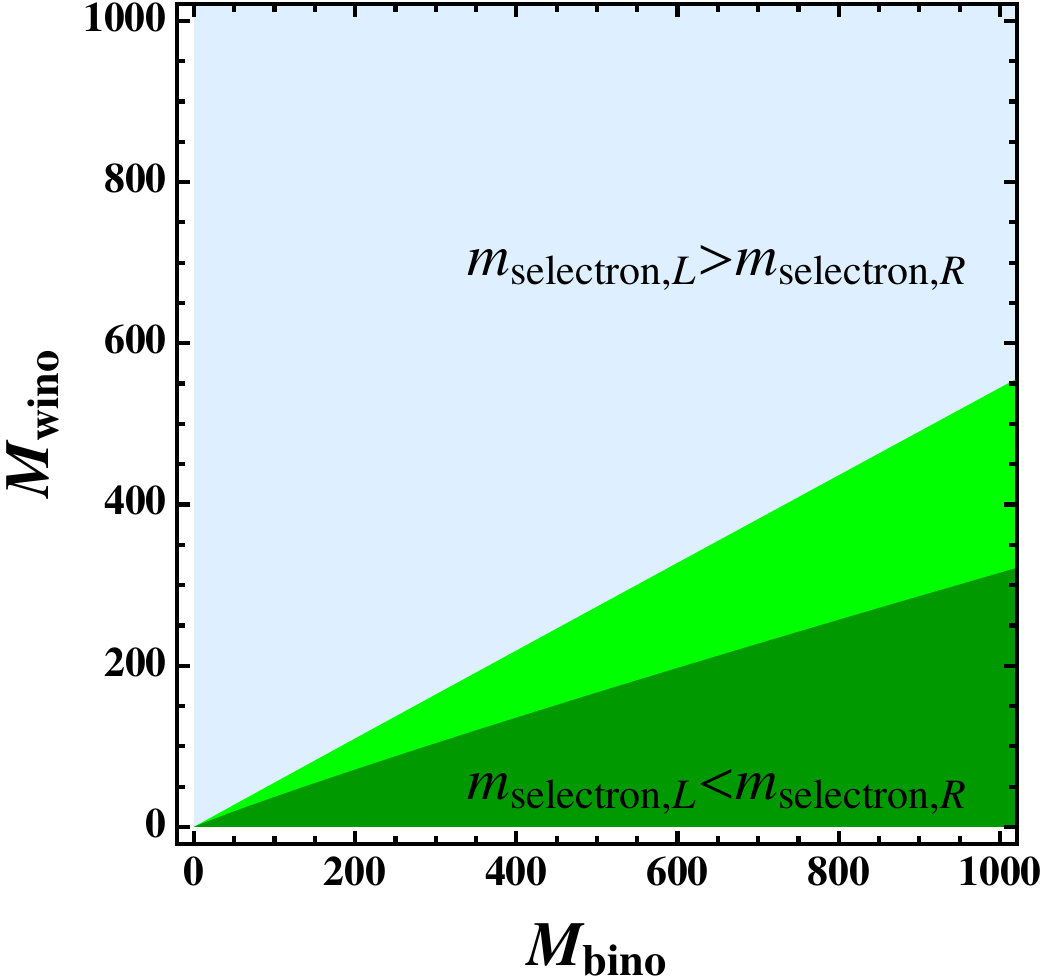}
\caption{Left and right handed selectron masses for a non-chiral input at the
  messenger scale, as well as $S=0$.  In the blue (dark green) region the left (right)
  handed selectrons are heavier for $\med > 10^5$~GeV. In the light green
  region the right handed selectrons can be heavier for sufficiently large
  $\med$.}
\label{fig:ordering}
\end{figure}

Our discussion so far has been in terms of bino and wino components of
the electroweak gauginos, but as already noted, due to the effects of
electroweak symmetry breaking the bino and wino are not the
appropriate mass eigenstates. Their mass matrix is given by
\begin{equation}
M_{\nz{}}=
\left(
\begin{array}{cccc}
M_1  & 0  & -c_\beta s_w m_Z & s_\beta s_w m_Z  \\
 0 & M_2  & c_\beta c_w m_Z & -s_\beta c_w m_Z \\
  -c_\beta s_w m_Z& c_\beta c_w m_Z  &0 &-\mu   \\
s_\beta s_w m_Z & -s_\beta c_w m_Z & -\mu & 0
\end{array}
\right),
\label{eq:neutralinomatrix}
\end{equation}
where $s_w=\sin \theta_w$, $c_w=\cos \theta_w$, etc.  This mass matrix
is real and symmetric, so its eigenvalues are real. Accordingly, the
mass matrix squared is positive definite and its smallest eigenvalue
is smaller than any of its diagonal elements
\begin{alignat}{5}
\min \, \mne{}
&< \min
\left[ \sqrt{M_1^2 + m_Z^2 \frac{1-\cos(2\theta_w)}{2}},
       \sqrt{M_2^2 + m_Z^2 \frac{1+\cos(2\theta_w)}{2}},
       \sqrt{\mu^2 + m_Z^2 \frac{1\pm\cos(2\beta)}{2}}
\right]
\notag \\
&< \min
\left[ \sqrt{M_1^2 + m_Z^2 \frac{1-\cos(2\theta_w)}{2}},
       \sqrt{M_2^2 + m_Z^2 \frac{1+\cos(2\theta_w)}{2}}
\right] \; .
\end{alignat}
For $M_{1,2} \gg m_Z$ the smallest eigenvalue of the neutralino mass
matrix is usually smaller than both $M_1$ and $M_2$.  Therefore, the
minimum curves for $\mse{}/M_1$ in Fig.~\ref{fig:mse_slope} also set a
lower limit on the ratio of selectron/slepton to the smallest
neutralino mass.

Because of the wealth of additional parameters the relevant question
is if these bounds are saturated. In a first attempt we assume gaugino
mass unification, which means the bino is roughly 6 times lighter than
the gluino. Current LHC constraints imply $\mgo > 750$~GeV,
translating into $M_1 > 125$~GeV, so our original assumption $M_1\gg
m_Z$ is reasonable. For illustration purposes we also assume large $\mu$, so
we can consider the limit $m_Z \ll
|M_1\pm\mu|,|M_2\pm\mu|$ and $M_{1,2} \ll \mu$.  In this regime the
lightest neutralino is bino-like and its mass is given 
\begin{equation}
\mne{1} = M_1 - \frac{m_Z^2  s_w^2}{\mu^2-M_1^2} \,
\left[ M_1 + \mu s_{2\beta} \right]
= M_1 \left[1+ \ord\left( \frac{m_Z^2}{\mu M_{1}} \right)\right] \; .
\label{eq:approx}
\end{equation}
In this limit the bound in the slepton-gaugino mass plane can indeed
be saturated.  
Our expectations for the ratios between neutralino and bino masses 
are confirmed in the test models briefly
discussed in the next section. Corresponding points are shown in the
left panel of Fig.~\ref{fig:dotted}.

One might be curious to see how the SUSY benchmark points included in
the squark-gluino plane in Fig.~\ref{fig:sq_go} are distributed on the
electroweak mass planes.  The black dots in the left panel of
Fig.~\ref{fig:mse_plane} denote values of the mass for
$(\mne{1},\mse{R})$ for those benchmark points.  As expected, they lie
in the high-scale region.  The only point located in the green region
corresponding to messenger scales below $10^{16}$~GeV is a gauge
mediated point with a very low messenger scale of 80~TeV.  We will
continue the discussion of benchmark points and models in the next
section.

In the left panel of Fig.~\ref{fig:mse_plane} we also introduce a ``No
Neutralino Dark Matter'' line.  Below it, a bino-like neutralino
cannot be dark matter.  It would decay into the lighter right handed
selectron which cannot be dark matter, as it is charged.  This
requirement is strongly correlated with a high mediation scale, \ie
once we require the bino to be the dark matter candidate we
automatically constrain the available parameter space to the fraction
accessible by high scale SUSY breaking.  Perhaps an obvious point to
note is that if dark matter is not made of neutralinos all points
below the dashed line remain perfectly viable.

If the lightest neutralino is bino-like in the limit of large $|\mu|$,
the second lightest neutralino is wino-like. In this case we can
interpret dark green region in the right hand panel of
Fig.~\ref{fig:mse_plane} as the area for $\nz{2}$ vs. $\mse{L}$
inaccessible to high scale models. However, here we need to be careful
with possible gaugino-Higgsino mixing effects.\bigskip

We can apply the same argument as for neutralinos to the chargino sector
with its mass matrix
\begin{equation}
M_{\cpm{}}=
\left(
\begin{array}{cc}
 0   & X^{T}  \\
 X & 0
\end{array}
\right)
\qquad
\text{with}
\quad
X=
\left(
\begin{array}{cc}
 M_2 & \sqrt 2 s_\beta m_W   \\
 \sqrt 2 c_\beta m_W & \mu
\end{array}
\right) \; .
\label{eq:charginomass1}
\end{equation}
Its eigenvalues are given by (each twice),
\begin{equation}
m^2_{\cpm{j}}
= \frac{1}{2}
\left[ |M_2|^2 + |\mu|^2 + 2m_W^2
      \mp \sqrt{ (|M_2|^2+|\mu|^2+2m_W^2)^2
                -4|\mu M_2-m_W^2 s_{2\beta}^2|^2}
\right] \; .
\end{equation}
Again, we find that the smallest eigenvalue is bounded from above as
\begin{equation}
\min \, m_{\cpm{}} <
\min \left[\sqrt{M_2^2+2 s_\beta^2 m_W^2},
                 \sqrt{M_2^2+2 c_\beta^2 m_W^2}
           \right] <
\sqrt{M_2^2+m_W^2}.
\end{equation}
For $m_W\ll M_2$ this smallest eigenvalue is typically below
$M_2$. Therefore, the wino curves for $\mse{}/M_2$ in
Fig.~\ref{fig:mse_slope} can be interpreted as lower limits on
selectron to lightest chargino mass ratio as a function of the
messenger mass.
Hence, the separation into three regions in the right hand panel of Fig.~\ref{fig:mse_plane} can 
be directly interpreted in terms of physical masses.
Again, for illustration we have indicated the distribution of the benchmark points.

\section{Benchmark and test models}
\label{sec:bench}

As shown in Section~\ref{sec:msq}, no supersymmetric model arising
from a theory with a high scale and containing Majorana gauginos can
cover the squark-gluino mass plane.  As a result, large regions in the
squark-gluino and slepton-gaugino mass planes are not populated by
such high scale models.  For example, assuming gravity mediation at
$\mgut$, only roughly half of the parameter space corresponding to
the light blue area in Fig.~\ref{fig:sq_go} will be covered. For gauge
mediation this effect is slightly more moderate as such models can
enter into and (if the mediation scale is chosen suitably low) cover
the green area in Fig.~\ref{fig:sq_go}.\bigskip

This shortcoming becomes particularly obvious when we study benchmark
points provided by theorists to help guide the LHC experiments.
Ref.~\cite{first} lists a standard set of the benchmark compiled for
and used at the LHC~\cite{atlas,cms,sps,adjk2}.  These benchmarks are
shown as black dots in Fig.~\ref{fig:sq_go}.  The first and most
important requirement on benchmark points is to represent the
available parameter space. The distribution we observe in
Fig.~\ref{fig:sq_go} clearly shows that this is not the case, provided
we consider the weak-scale MSSM the model the LHC looks for.  All
benchmark points populate the region of the squark-gluino mass plane
which can be linked to high-scale SUSY breaking. In addition,
reminiscent of the population of Scotland (or Canada), the vast majority of
benchmark points in Fig.~\ref{fig:sq_go} live along the southern
border which saturates the $\msq{}/\mgo$ mass ratio, \ie values of the
squark mass where the renormalization group induced contribution shown
in Eq.\eqref{eq:slope} dominates over the soft breaking scalar
mass. One of the underlying reasons for this squeezed distribution is
that most of the benchmark points are CMSSM points. In the CMSSM all
sfermions have the same initial mass at the GUT scale characterized by
the parameter $m_{0}$ which is typically chosen to be of the order of
the electroweak scale.  At the same time the contributions arising
from gauginos scale with their gauge couplings and gluino
contributions are therefore dominant.\bigskip

For the weakly interacting particles all but one benchmark point also
lie in the upper region.  Indeed by construction all those points lie
even above the ``No Neutralino Dark Matter'' line.  The benchmark
points are now more spread out because the initial value of the
universal CMSSM sfermion mass is comparable to electroweak gaugino
contributions.  Nevertheless, they still cover only a restricted area
of parameter space.

\begin{figure}[t]
\includegraphics[width=0.40\textwidth]{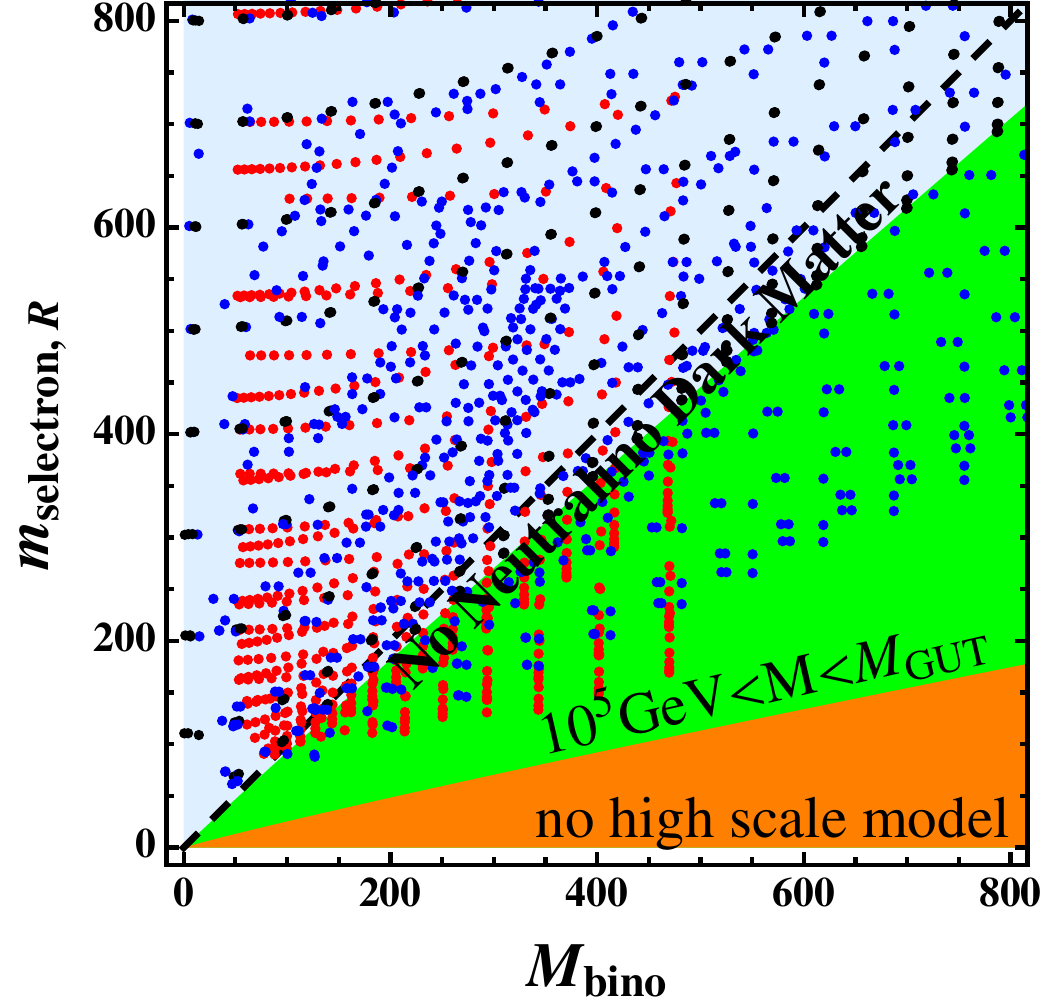}
\hspace{0.1\textwidth}
\includegraphics[width=0.40\textwidth]{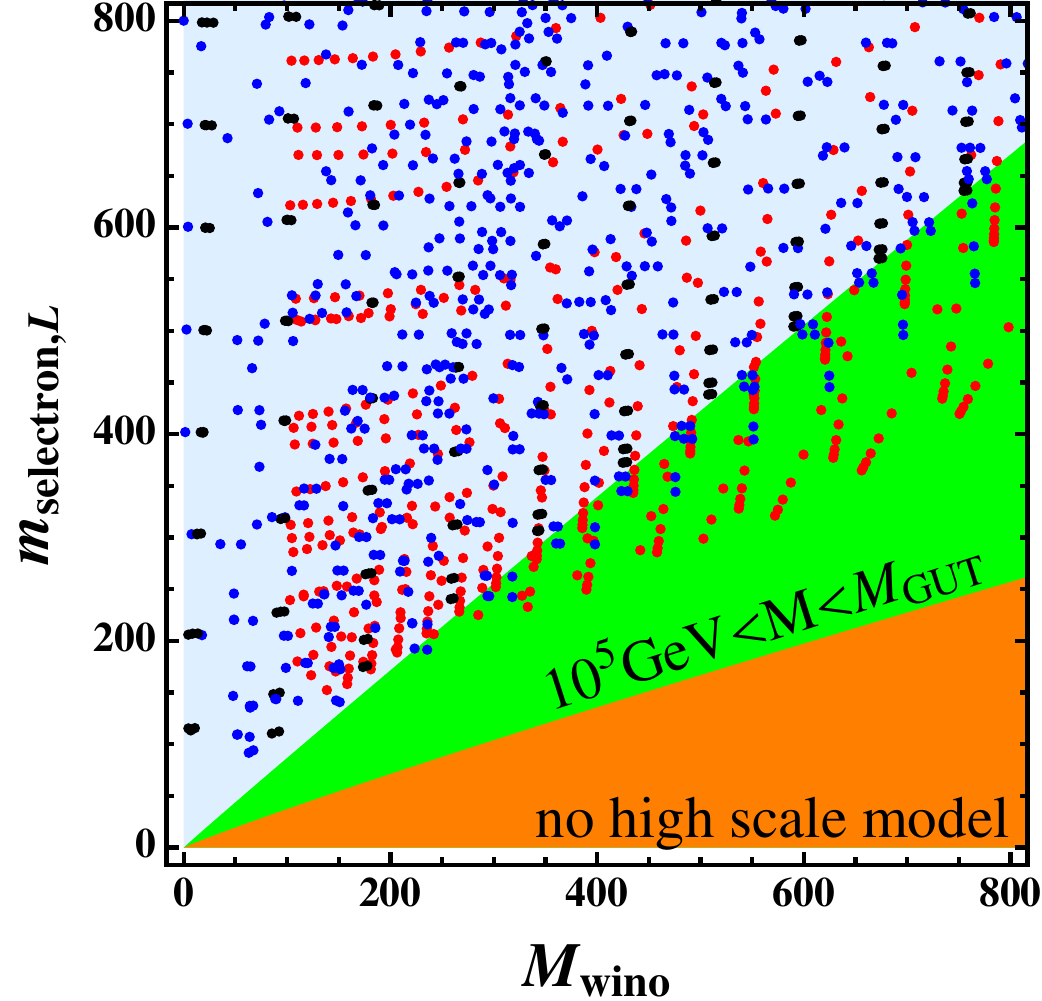}
\caption{The sfermion-gaugino mass plane, for a bino or wino and left
  and right selectrons and with the same color coding as
  Fig.~\ref{fig:sq_go}.  The dots represent scans over high-scale
  models, namely the CMSSM (black), a low-scale CMSSM (blue) and pure
  generalized gauge mediation (red) (see text for details).}
\label{fig:dotted}
\end{figure}

To illustrate more generally (\ie not just based on the limited set of
benchmark points) how the sfermion-gaugino mass plane is populated by
high-scale models we show in Fig.~\ref{fig:dotted} a large set of
parameter points scanning over a variety of test models:
\begin{itemize}
\item[--] the CMSSM with $\tan \beta =3,10,40$ and $A_0=0$;
\item[--] the same initial soft parameters ($\tan \beta =3,10,40$,
  $A_0=0$) but at lower $\med=2 \cdot 10^6, 2 \cdot 10^{10}$~GeV;
\item[--] pure general gauge mediation (pGGM) with
  $\mess=10^8,10^{10},10^{14}$~GeV, see Refs.~\cite{adjk2,adjk1,first} for
  details.
\end{itemize}

Following our discussion in the previous section we use for the $x$
axis coordinates the masses of the lightest neutralino $\mne{1}$ and
the lightest chargino $\mch{1}$.  The (black) CMSSM points indeed
cover the accessible parameter space and saturate the minimal ratios
for $\mse{}/M_{1,2}$.  The (blue) low-scale ``gravity mediation''
points extend into the intermediate $\med$ wedge though they do not
approach the lower end it.  The pure general gauge mediation points
marked in red extend further into this intermediate region.  The same
models have a qualitatively very similar behavior on the squark-gluino
mass plane.

We also note that in the left panel the pGGM points do not extend to
arbitrarily high neutralino masses.  This is special to this model
which becomes non-perturbative for parameter values that correspond to
large bino masses.

\section{Conclusions}
\label{sec:conclusions}

Many LHC searches for supersymmetry are conveniently interpreted on the
squark-gluino mass plane.  In this note we have argued that in the
MSSM all sfermion-gaugino mass planes can be divided into three wedge
shaped-regions: the first region with high squark masses is accessible to
all types of SUSY models including those with a high mediation scale
$M\gtrsim M_{\rm GUT}$. The second region with intermediate values of
the sfermion to gaugino mass ratio requires a mediation scale $M<M_{\rm
  GUT}$. Finally, the third region with the low sfermion to gaugino
mass ratios cannot be accessed by any MSSM type model with a mediation
scale $M\geq 10^{5}\,{\rm GeV}$.  The models in this third wedge would
have to be described by an exotic SUSY theory.  Discovering SUSY in
this region would be a particularly surprising and exciting
outcome.\bigskip

What does ``exotic'' mean in this context and how might such theories
look? In general, any renormalization group evolution of scalar masses
sufficiently different from the one considered here could result in a
theory living in the third wedge. The renormalization group equations
we employed are inherently MSSM equations. A non-MSSM matter content
could therefore give an exotic theory. One well understood example of
this are models with Dirac gauginos~\cite{dirac}. In these theories the
Dirac gaugino masses simply do not determine the running the sfermion
masses~\cite{dirac2}. Of course, this is just one example of an exotic theory
arising from a non-MSSM setup.

One of our technical assumptions was that the sfermion masses at the
mediation should not be non-tachyonic. In principle, allowing such
tachyons is a way to lower the physical sfermion to gaugino mass
ratios below minimal values for high scale models we have computed in
this note.  The examples of these models examined in~\cite{tachyons} often contain low lying color breaking vacua and while in general models of this type are not necessarily excluded they need to be carefully screened for dangerous instabilities.

A third large class of exotic theories are models without a
significant separation between the electroweak scale and the scale at
which the soft terms are generated. Practically, such models could be
described by effective actions with soft terms defined at the collider
scale, avoiding any renormalization group evolution. An even simpler
approach to generate model points would be to use various versions of
simplified models~\cite{simplified}.

From an LHC perspective the striking result of our study is that these
fundamentally very different structures can be classified in terms of
the standard scalar-gaugino mass planes and that their physics is
essentially determined by one parameter, the mediation scale of SUSY
breaking.

\acknowledgments VVK is grateful to the Aspen Center for Physics where
part of this work was carried out. TP would like to thank Tao Han for
his hospitality in Madison, where part of this text was typed on and
off the Union Terrace.


\begin{thebibliography}{10}
\bibitem{atlas}
 G.~Aad {\it et al.}  [The ATLAS Collaboration],
  arXiv:0901.0512 [hep-ex];
 J.~B.~G.~da Costa {\it et al.}  [Atlas Collaboration],
  Phys.\ Lett.\  B {\bf 701} (2011) 186
  [arXiv:1102.5290 [hep-ex]];
 D.~Charlton for the ATLAS collaboration,
  {\it Highlights and Searches in ATLAS},
  Talk at the Europhysics conference on high energy physics, Grenoble, 2011;
 W.~Ehrenfeld for the ATLAS collaboration,
  {\it SUSY Searches in ATLAS},
  Talk at SUSY, Fermilab, 2011.

\bibitem{cms}
 G.~L.~Bayatian {\it et al.}  [CMS Collaboration],
  J.\ Phys.\ G {\bf 34}, 995 (2007);
 CMS Collaboration, 
  CMS PAS SUS-10-005, 2011;
 G.~Tonelli for the CMS collaboration,
  {\it Highlights and Searches in CMS},
  Talk at the Europhysics conference on high energy physics, Grenoble, 2011.
 I.~Melzer-Pellmann for the CMS collaboration,
  {\it SUSY Searches in CMS},
  Talk at SUSY, Fermilab, 2011.

\bibitem{flavor}
 L.~J.~Hall, V.~A.~Kostelecky and S.~Raby,
  Nucl.\ Phys.\  B {\bf 267}, 415 (1986);
 G.~D'Ambrosio, G.~F.~Giudice, G.~Isidori and A.~Strumia,
 Nucl.\ Phys.\  B {\bf 645}, 155 (2002);
  [arXiv:hep-ph/0207036].
 for a recent compilation of constraints see \eg 
  S.~Dittmaier, G.~Hiller, T.~Plehn and M.~Spannowsky,
  arXiv:0708.0940 [hep-ph].

\bibitem{edm}
 V.~D.~Barger, T.~Falk, T.~Han, J.~Jiang, T.~Li, T.~Plehn,
  Phys.\ Rev.\  {\bf D64}, 056007 (2001)
  [hep-ph/0101106];
 S.~Abel, S.~Khalil, O.~Lebedev,
  Nucl.\ Phys.\  {\bf B606}, 151-182 (2001)
  [hep-ph/0103320];
 M.~Pospelov, A.~Ritz,
  Annals Phys.\  {\bf 318}, 119-169 (2005)
  [hep-ph/0504231].

\bibitem{sfitter}
 G.~A.~Blair, W.~Porod, P.~M.~Zerwas,
  Phys.\ Rev.\  {\bf D63}, 017703 (2001)
  [hep-ph/0007107];
 P.~Bechtle, K.~Desch, P.~Wienemann,
  Comput.\ Phys.\ Commun.\  {\bf 174}, 47-70 (2006)
  [hep-ph/0412012];
 R.~Lafaye, T.~Plehn, M.~Rauch, D.~Zerwas,
  Eur.\ Phys.\ J.\  {\bf C54}, 617-644 (2008)
  [arXiv:0709.3985 [hep-ph]];
 S.~S.~AbdusSalam, B.~C.~Allanach, F.~Quevedo, F.~Feroz and M.~Hobson,
  Phys.\ Rev.\  D {\bf 81} (2010) 095012
  [arXiv:0904.2548 [hep-ph]].
 C.~Adam, J.~-L.~Kneur, R.~Lafaye, T.~Plehn, M.~Rauch, D.~Zerwas,
  Eur.\ Phys.\ J.\  {\bf C71}, 1520 (2011)
  [arXiv:1007.2190 [hep-ph]].  

\bibitem{Chamseddine:1982jx}
  A.~H.~Chamseddine, R.~L.~Arnowitt and P.~Nath,
  Phys.\ Rev.\ Lett.\  {\bf 49} (1982) 970;
  R.~Barbieri, S.~Ferrara and C.~A.~Savoy,
  Phys.\ Lett.\  B {\bf 119} (1982) 343;
  L.~J.~Hall, J.~D.~Lykken and S.~Weinberg,
  Phys.\ Rev.\  D {\bf 27} (1983) 2359;
  H.~P.~Nilles,
  Phys.\ Rept.\  {\bf 110} (1984) 1.


\bibitem{Tata:1997uf}
 for a review see X.~Tata,
  arXiv:hep-ph/9706307.

\bibitem{Dine:1981gu}
  M.~Dine and W.~Fischler,
  Phys.\ Lett.\  B {\bf 110} (1982) 227;
  C.~R.~Nappi and B.~A.~Ovrut,
  Phys.\ Lett.\  B {\bf 113} (1982) 175;
  L.~Alvarez-Gaume, M.~Claudson and M.~B.~Wise,
  Nucl.\ Phys.\  B {\bf 207} (1982) 96;
  M.~Dine, A.~E.~Nelson and Y.~Shirman,
  Phys.\ Rev.\  D {\bf 51} (1995) 1362
  [arXiv:hep-ph/9408384].

\bibitem{Meade:2008wd}
  P.~Meade, N.~Seiberg and D.~Shih,
  Prog.\ Theor.\ Phys.\ Suppl.\  {\bf 177} (2009) 143
  [arXiv:0801.3278 [hep-ph]].

\bibitem{Giudice:1998bp}
for a review see  G.~F.~Giudice and R.~Rattazzi,
  Phys.\ Rept.\  {\bf 322} (1999) 419
  [arXiv:hep-ph/9801271].

\bibitem{Randall:1998uk}
  L.~Randall and R.~Sundrum,
  Nucl.\ Phys.\  B {\bf 557} (1999) 79
  [arXiv:hep-th/9810155];
  G.~F.~Giudice, M.~A.~Luty, H.~Murayama and R.~Rattazzi,
  JHEP {\bf 9812} (1998) 027
  [arXiv:hep-ph/9810442].

\bibitem{primer}
 M.~Drees, S.~P.~Martin,
  [hep-ph/9504324];
 S.~P.~Martin,
  [hep-ph/9709356].

\bibitem{susy_rge}
 B.~C.~Allanach,
  Comput.\ Phys.\ Commun.\  {\bf 143}, 305-331 (2002)
  [hep-ph/0104145];
 A.~Djouadi, J.~-L.~Kneur, G.~Moultaka,
  Comput.\ Phys.\ Commun.\  {\bf 176}, 426-455 (2007)
  [hep-ph/0211331];
 W.~Porod,
  Comput.\ Phys.\ Commun.\  {\bf 153}, 275-315 (2003)
  [hep-ph/0301101];
 B.~C.~Allanach, S.~Kraml, W.~Porod,
  JHEP {\bf 0303}, 016 (2003)
  [hep-ph/0302102];
 I.~Jack, D.~R.~T.~Jones, A.~F.~Kord,
  Phys.\ Lett.\  {\bf B579}, 180-188 (2004)
  [hep-ph/0308231];
 S.~P.~Martin,
  Phys.\ Rev.\  {\bf D71}, 116004 (2005)
  [hep-ph/0502168].

\bibitem{pmssm}
  A.~Djouadi {\it et al.}  [MSSM Working Group],
  arXiv:hep-ph/9901246.
  

\bibitem{simplified}
  D.~Alves, N.~Arkani-Hamed, S.~Arora, Y.~Bai, M.~Baumgart, J.~Berger, M.~Buckley, B.~Butler {\it et al.},
  [arXiv:1105.2838 [hep-ph]].
  


\bibitem{split_susy}
 N.~Arkani-Hamed, S.~Dimopoulos,
  JHEP {\bf 0506}, 073 (2005)
  [hep-th/0405159];
 G.~F.~Giudice, A.~Romanino,
  Nucl.\ Phys.\  {\bf B699}, 65-89 (2004)
  [hep-ph/0406088];
 W.~Kilian, T.~Plehn, P.~Richardson, E.~Schmidt,
  Eur.\ Phys.\ J.\  {\bf C39}, 229-243 (2005)
  [hep-ph/0408088];
 J.~D.~Wells,
  Phys.\ Rev.\  {\bf D71}, 015013 (2005)
  [hep-ph/0411041]
 D.~S.~M.~Alves, E.~Izaguirre, J.~G.~Wacker,
  [arXiv:1108.3390 [hep-ph]].

\bibitem{tachyons}
 J.~L.~Feng, A.~Rajaraman, B.~T.~Smith,
  Phys.\ Rev.\  {\bf D74}, 015013 (2006)
  [hep-ph/0512172].
 A.~Rajaraman, B.~T.~Smith,
  Phys.\ Rev.\  {\bf D75}, 115015 (2007)
  [hep-ph/0612235];
 J.~R.~Ellis, J.~Giedt, O.~Lebedev, K.~Olive, M.~Srednicki,
  Phys.\ Rev.\  {\bf D78}, 075006 (2008)
  [arXiv:0806.3648 [hep-ph]].

  \bibitem{ATLASsimpl}
 I.~Vivarelli for the ATLAS collaboration,
  {\it 	Search for supersymmetry in jet(s) plus missing transverse momentum final states with the ATLAS detector},
  Talk at the Europhysics conference on high energy physics, Grenoble, 2011;

\bibitem{sps}
  B.~C.~Allanach, M.~Battaglia, G.~A.~Blair, M.~S.~Carena, A.~De Roeck, A.~Dedes, A.~Djouadi, D.~Gerdes {\it et al.},
  Eur.\ Phys.\ J.\  {\bf C25}, 113-123 (2002)
  [hep-ph/0202233].

\bibitem{adjk2}
  S.~Abel, M.~J.~Dolan, J.~Jaeckel and V.~V.~Khoze,
  JHEP {\bf 1012} (2010) 049
  [arXiv:1009.1164 [hep-ph]].

\bibitem{how_light}
 D.~Hooper, T.~Plehn,
  Phys.\ Lett.\  {\bf B562}, 18-27 (2003)
  [hep-ph/0212226].
  H.~K.~Dreiner, S.~Heinemeyer, O.~Kittel, U.~Langenfeld, A.~M.~Weber, G.~Weiglein,
  Eur.\ Phys.\ J.\  {\bf C62}, 547-572 (2009)
  [arXiv:0901.3485 [hep-ph]].

\bibitem{jet_rad}
  T.~Plehn, D.~Rainwater, P.~Z.~Skands,
  Phys.\ Lett.\  {\bf B645}, 217-221 (2007)
  [hep-ph/0510144];
 T.~Plehn, T.~M.~P.~Tait,
  J.\ Phys.\ G {\bf G36}, 075001 (2009)
  [arXiv:0810.3919 [hep-ph]];
 J.~Alwall, S.~de Visscher, F.~Maltoni,
  JHEP {\bf 0902}, 017 (2009)
  [arXiv:0810.5350 [hep-ph]].


\bibitem{autofocus}
  C.~Englert, T.~Plehn, P.~Schichtel, S.~Schumann,
  Phys.\ Rev.\  {\bf D83}, 095009 (2011)
  [arXiv:1102.4615 [hep-ph]].

\bibitem{osland}
 B.~K.~Gjelsten, D.~J.~Miller, P.~Osland,
  JHEP {\bf 0506}, 015 (2005)
  [hep-ph/0501033].

\bibitem{adjk1}
  S.~Abel, M.~J.~Dolan, J.~Jaeckel and V.~V.~Khoze,
  JHEP {\bf 0912} (2009) 001
  [arXiv:0910.2674 [hep-ph]].

\bibitem{first}
  M.~J.~Dolan, D.~Grellscheid, J.~Jaeckel, V.~V.~Khoze and P.~Richardson,
  JHEP {\bf 1106} (2011) 095
  [arXiv:1104.0585 [hep-ph]].
  
\bibitem{dirac}
 L.~J.~Hall and L.~Randall,
  Nucl.\ Phys.\  B {\bf 352} (1991) 289;
 G.~D.~Kribs, E.~Poppitz and N.~Weiner,
  Phys.\ Rev.\  D {\bf 78} (2008) 055010
  [arXiv:0712.2039 [hep-ph]];
 K.~Benakli and M.~D.~Goodsell,
  Nucl.\ Phys.\  B {\bf 816} (2009) 185
  [arXiv:0811.4409 [hep-ph]];
  S.~Abel and M.~Goodsell,
  JHEP {\bf 1106} (2011) 064;
  [arXiv:1102.0014 [hep-th]].

\enlargethispage{0.5cm}
\bibitem{dirac2}
K.~Benakli and M.~D.~Goodsell,
  Nucl.\ Phys.\  B {\bf 840} (2010) 1
  [arXiv:1003.4957 [hep-ph]];
 K.~Benakli,
  arXiv:1106.1649 [hep-ph].

\end{thebibliography}
\end{document}